\documentclass[apl,reprint]{revtex4-1}
\usepackage{graphicx}

\begin{document}

\title{Multi-dimensional laser spectroscopy of exciton polaritons with spatial light modulators}
\author{P. Mai}
\author{B. Pressl}
\author{M. Sassermann}
\author{Z. V\"or\"os}
\author{G. Weihs}
\affiliation{Institut f\"ur Experimentalphysik, Universit\"at Innsbruck, Technikerstra\ss{}e 25, 6020 Innsbruck, Austria}
\author{C. Schneider}
\author{A. L\"offler}
\author{S. H\"ofling}
\author{A. Forchel}
\affiliation{Technische Physik, Physikalisches Institut, Wilhelm Conrad R\"ontgen Research Center for Complex Material Systems, Universit\"at W\"urzburg, Am Hubland, D-97074 W\"urzburg, Germany}

\begin{abstract}
We describe an experimental system that allows one to easily access the dispersion curve of exciton-polaritons in a microcavity. Our approach is based on two spatial light modulators (SLM), one for changing the excitation angles (momenta), and the other for tuning the excitation wavelength. We show that with this setup, an arbitrary number of states can be excited accurately and that re-configuration of the excitation scheme can be done at high speed.
\end{abstract}

\maketitle

Cavity polaritons are the coherent superpositions of a confined electric field, and some excitation of the solid, be it a phonon, plasmon, or exciton. Amongst these quasi-particles, owing to both possible applications and their role in fundamental research, exciton-polaritons in a planar microcavity are perhaps the most extensively studied.

These quasi-particles are relatively short-lived (their lifetime is of the order of several ps), and they can decay by the emission of a photon. From the experimental point of view, one of the advantages of using a planar structure is that the momentum of the decaying polariton is mapped into the angle of the out-going photon. Conversely, by exciting the system at a particular angle and energy, a well-defined polariton state can be populated at will. While this seems simple in theory, practice tends to be less straightforward.

In order to excite, or to measure a specific polariton state, several approaches have been developed through the years. These involve either one or more goniometers \cite{Savvidis2000,Balili2009a,Kundermann2004}, a confocal setup \cite{Langbein2005,Savasta2005}, or beam shifting \cite{Huang2000}. Common to all these schemes is a mechanical means of moving the beams, which implies that re-configuration of the excitation geometry is slow and complicated. Furthermore, the complexity of the setup increases considerably, if more than one excitation beam is involved, and experiments become impractical for three beams.

In this Letter, we would like to introduce a scheme that does not suffer from the above-mentioned difficulties. Instead of mechanically moving optical elements, we apply spatial light modulators (SLM), thereby eliminating the main limiting factor. SLMs are special reflective or transmissive liquid crystal devices that can modulate either the phase, or amplitude of the impinging light, or both at the same time.

SLMs have found many uses, both in fundamental, and applied research. They are extensively utilized as optical tweezers in biological applications \cite{ChristianMaurer2007, Grier2003}, to beat the diffraction limit in optical imaging \cite{Putten2011}, or to create light with orbital angular momentum in quantum information experiments \cite{AlipashaVaziri2002, Jack2009}, just to name a few examples. This latter approach has successfully been applied in polariton experiments to create a polariton superfluid with a well-defined angular momentum \cite{Krizhanovskii2010,Lagougadis}.

\begin{figure}[t]
 \includegraphics[width=0.47\textwidth]{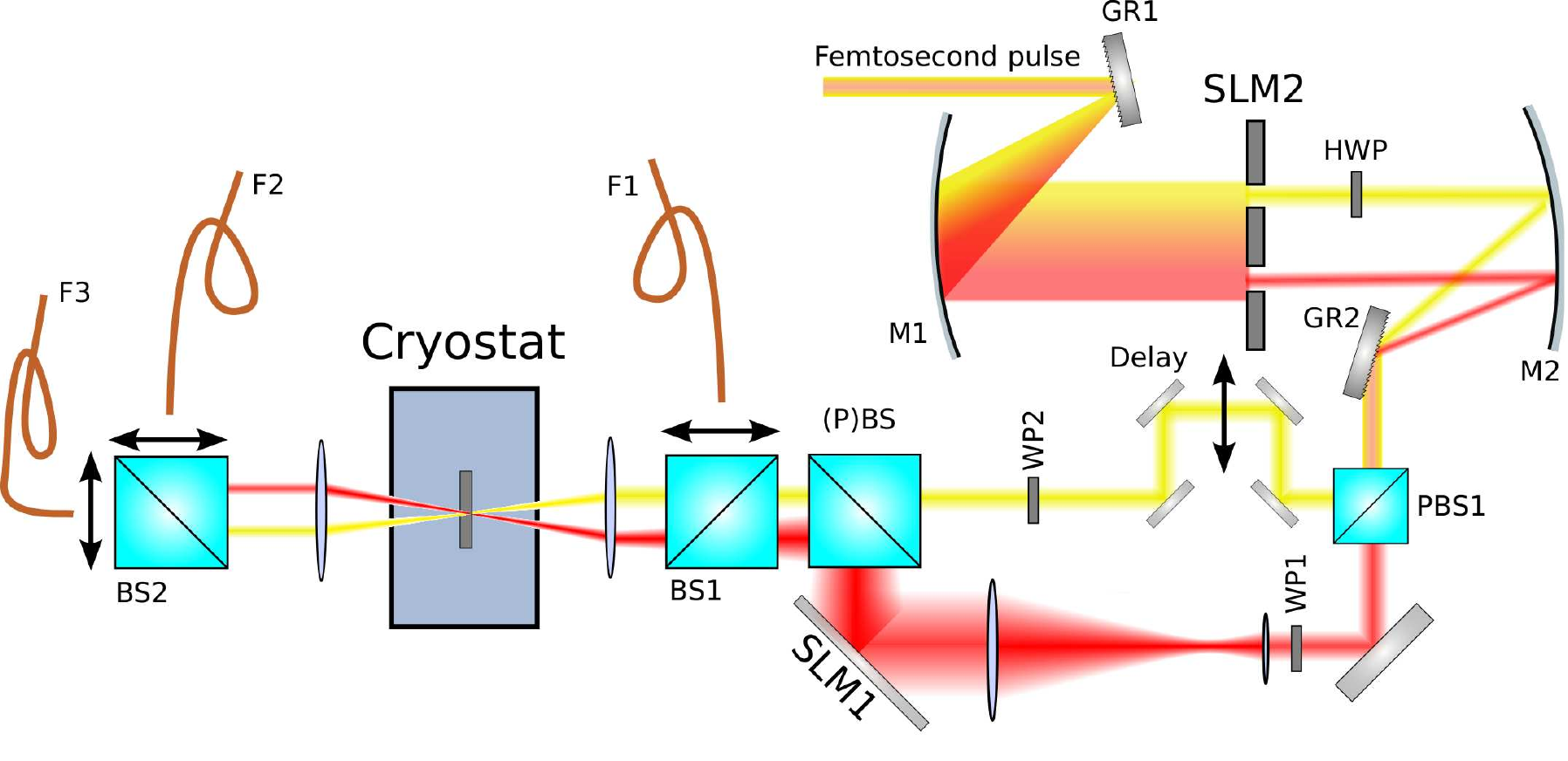}
 \caption{(color online) Experimental configuration: A phase-only SLM (SLM1) is used to displace the beam, thus to change the excitation momentum. SLM2 is an amplitude-only device, which selects the excitation wavelengths in a standard \protect{4-\textit{f}} pulse compressor. The half-wave plate (HWP) and the polarizing beam splitter (PBS1) are optional, and are used only when two-color excitation is required. The optical fibers are used in momentum-selective detection, either in reflection (F1), or transmission configuration (F2, F3). For clarity, not all split beams are shown.}
 \label{fig:exp_setup}
\end{figure}

\begin{figure}[t]
 \includegraphics[width=0.47\textwidth]{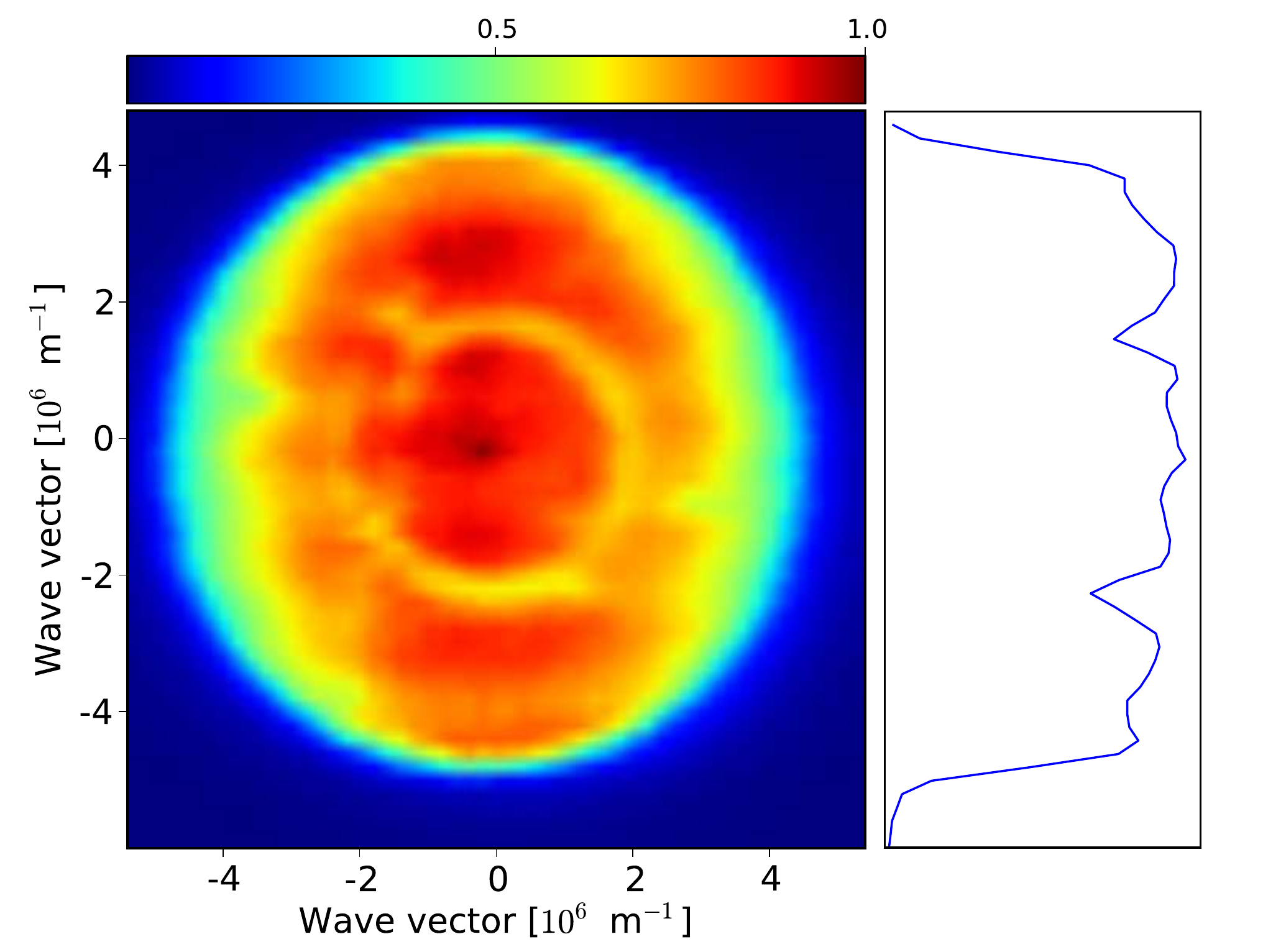}
 \caption{(color online) Reflectivity scan of an AlGaAs/GaAs microcavity. The color range is relative to the maximum of the reflected intensity. The intensity cut-off at around \protect{$4.5\cdot 10^6$\,m${}^{-1}$} is due to the numerical aperture of the microscope objective. The right hand side shows the intensity along the vertical line passing through $k=0$.}
 \label{fig:reflectivity_scan}
\end{figure}

\begin{figure}
 \includegraphics[width=0.47\textwidth]{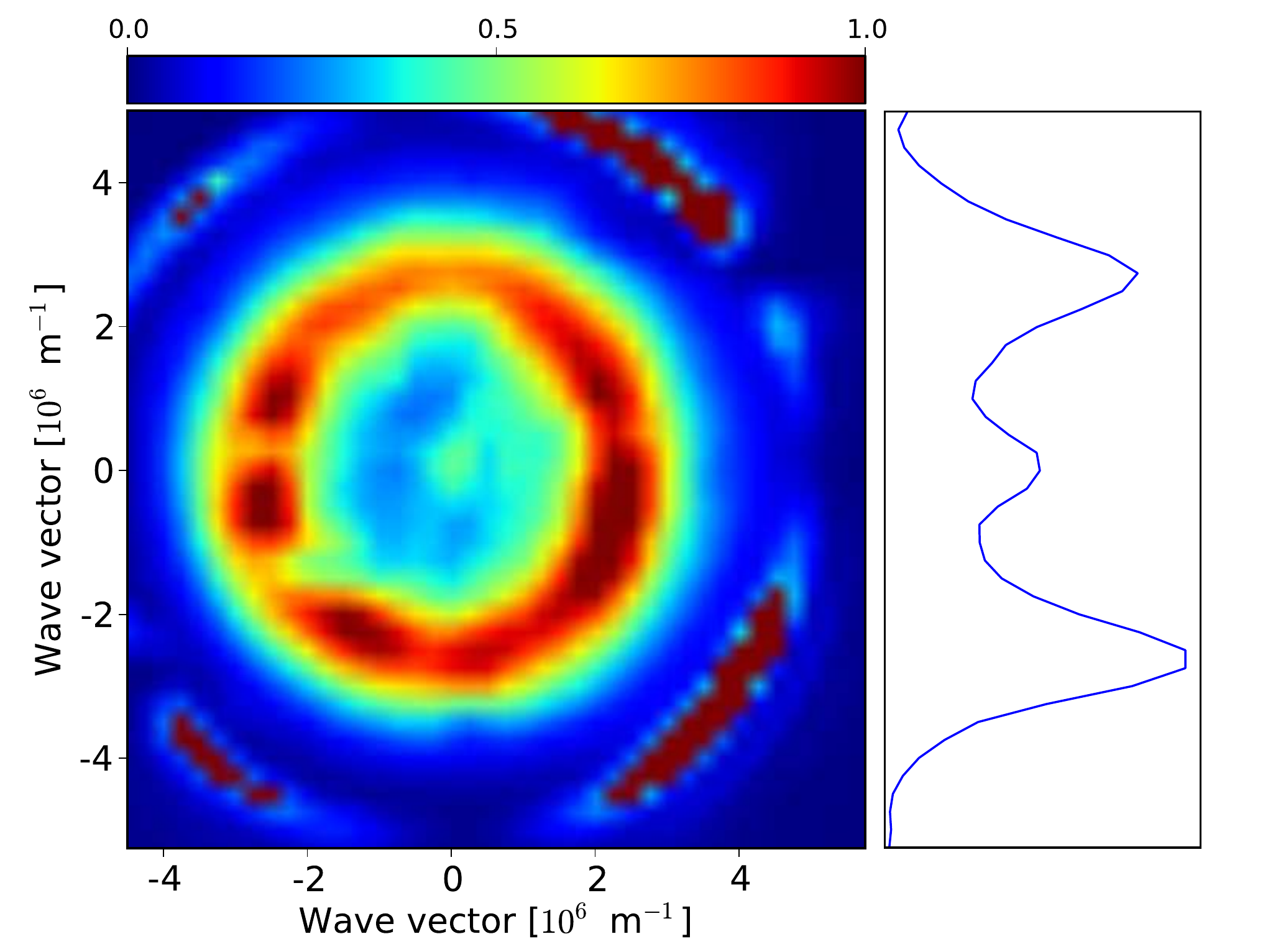}
 \caption{(color online) Luminescence intensity at $k=0$ as a function of the excitation momentum. The color range is relative to the maximum of the luminescence intensity. On the right, a cut is shown along a vertical line through $k=0$. The red circle with a radius of about $k = 4.5\cdot 10^6$\,m${}^{-1}$ is laser light scattered by the aperture of the microscope objective, while the small peak at $k=0$ is the tail of the excitation laser's spectrum.}
 \label{fig:ple-combined}
\end{figure}

Our experimental setup is shown in Fig.~\ref{fig:exp_setup}. A collimated laser beam impinges on a phase-only spatial light modulator (SLM1) \cite{holoeye}. The beam expander only serves to make the illumination more uniform. On the SLM, we define a phase-mask equivalent to a lens (a parabolic phase profile), thereby focusing the beam to a 100-$\mu$m large spot on the entrance side of a microscope objective with working distance of 2.3\,mm \cite{zeiss}. It is important to note that since the beam is focused on the objective, from the objective's perspective, the incoming beam can be approximated as a localized plane wave, which is, in turn, focused by the objective to an about 10-$\mu$m large spot on the sample. By moving the SLM's phase profile, the focal point can be shifted in the plane of the microscope objective, thus changing the excitation angle. With the 100-$\mu$m focal spot on the objective, we can achieve an angular selection of about 2 degrees. This translates into about $2\cdot 10^5$\,m${}^{-1}$ momentum selection in the total accessible polariton momentum range of about $5\cdot 10^6$\,m${}^{-1}$ \cite{note}. (In our case, the accessible momentum range is limited by the clear aperture of the objective, shown in Figs.~\ref{fig:reflectivity_scan}, and \ref{fig:ple-combined}.)

When tuning of the laser's energy is required, we make the laser pass through a standard pulse-shaping configuration containing two 1200-line gratings, and a pair of spherical mirrors with a focal length of 500 mm (upper half of Fig.~\ref{fig:exp_setup}). A laser beam with duration 180 fs is dispersed by the first grating (GR1), and then combined again with the second grating (GR2). In the common focal plane of the spherical mirrors (M1, M2), an amplitude-only SLM (SLM2) with 128 pixels is located \cite{cri}, and it acts as a re-configurable wavelength filter. The configuration above allows us to select an arbitrary wavelength within the bandwidth of laser pulse with a precision of about 0.1 nm. This setup can easily be extended, in order to make two-color excitation schemes possible: instead of one, two slits are defined on SLM2, and the polarization of one of the colors is rotated by 90 degrees by an achromatic waveplate (HWP). In this way, after re-combining the beams on GR2, we can separate them on a polarizing beam splitter (PBS), so that the two colors can be steered independently. The waveplates WP1, and WP2, and the delay line in one of the beams are optional, and can be used for generating arbitrary linear or circular polarization, and delays. We should point out that apart from translation stages required for adjusting the foci at the beginning of an experiment, the delay line is the only variable mechanical part in our setup.

In order to prove the viability of our setup, we carried out various measurements on a AlGaAs/GaAs microcavity sample identical to that discussed in \cite{Bloch1998}. The sample consists of a $\lambda/2$ cavity with three sets of four quantum wells located at the antinodes of the confined field.

\begin{figure}[b]
 \includegraphics[width=0.47\textwidth]{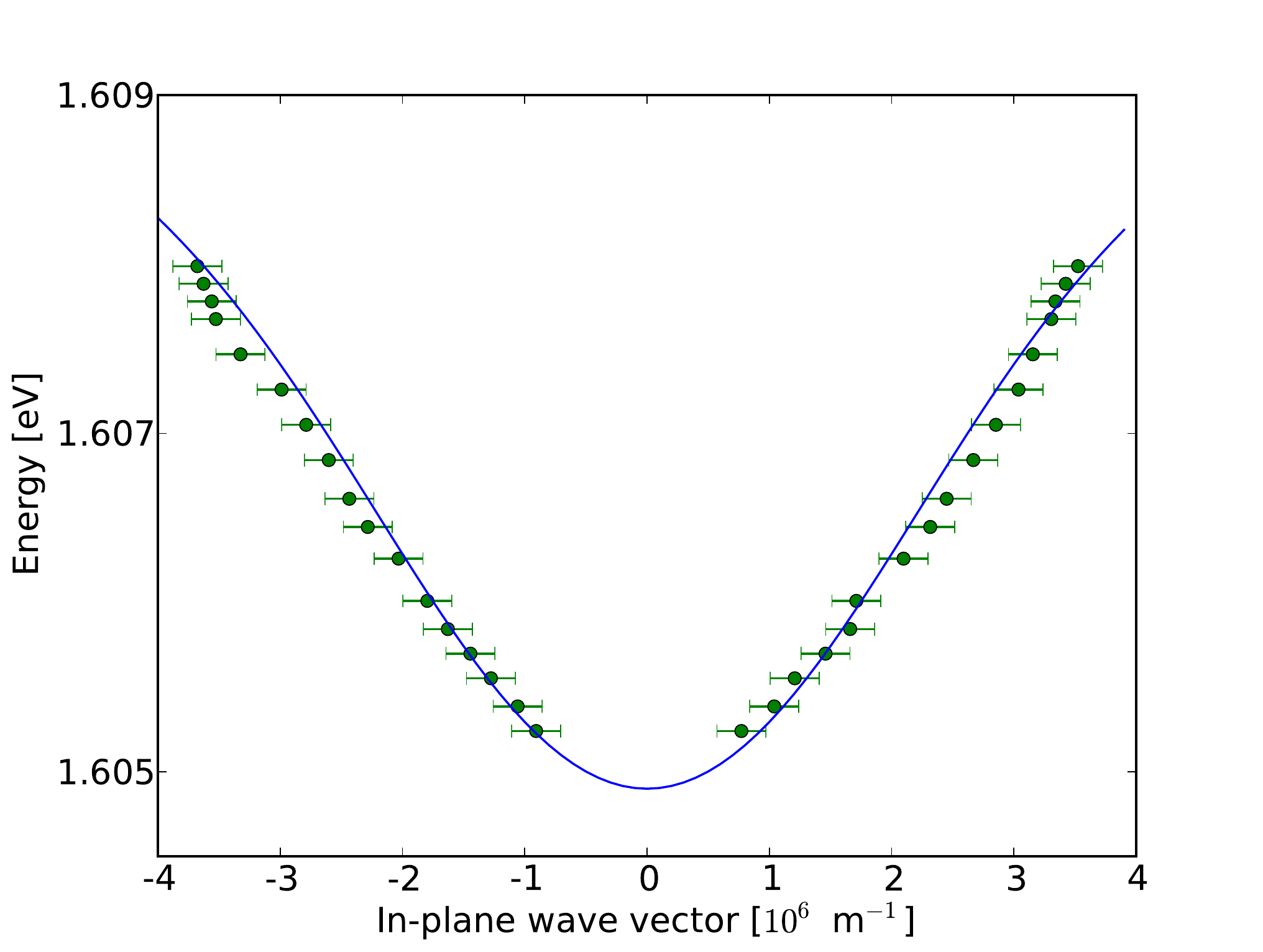}
 \caption{Lower polariton dispersion extracted from reflectivity scans similar to those in Fig.~\ref{fig:reflectivity_scan}. The solid line is a fit to the analytical lower polariton dispersion with parameters $\hbar\Omega_{\mathrm{Rabi}} = 11$\,meV, $E_c = 1.612$\,meV, and $E_x = 1.6088$\,meV \cite{Deng2010}. The horizontal error was determined from the error of the fit to the circles on Fig.~\ref{fig:reflectivity_scan}, while the error in the energy of the laser is less than 0.05\,meV, the size of the dots in the figure.}
 \label{fig:reflectivity_scan_spec}
\end{figure}

First, we measured the reflectivity of such a microcavity sample as a function of the excitation angle. For this, we kept the laser wavelength fixed at $\lambda=770.5$\,nm. The laser beam is split by a beam splitter, (BS1) in Fig.~\ref{fig:exp_setup}. SLM1 focuses the  passing component on the objective, while the other half serves as power reference. The beam reflected by the sample is then directed towards a large-area power meter, which replaces the fiber (F1) in Fig.~\ref{fig:exp_setup}. If the incoming laser beam is now scanned in the $x-y$ plane (the plane of the back of the microscope objective), the reflected intensities will be reduced, if the corresponding momentum coincides with that dictated by the polariton dispersion, and the energy of the laser. This polariton resonance manifests itself as a circle of lower intensity in Fig.~\ref{fig:reflectivity_scan}. The right hand side of the figure shows the intensity along the vertical line that passes through the origin of momentum space, and it clearly demonstrates the two resonances at $k=\pm 2\cdot 10^6$\,m${}^{-1}$.

By shifting the mask positions on SLM2, the measurement can be repeated for other energies, giving a three-dimensional map of the polariton dispersion. A two-dimensional projection of the dispersion of the lower polariton branch is shown in Fig.~\ref{fig:reflectivity_scan_spec}.
The points were obtained by fitting a circular profile to the reflectivity minimum in figures similar to Fig.~\ref{fig:reflectivity_scan}, and then taking the circles' coordinates at the leftmost, and rightmost points, respectively. In Fig.~\ref{fig:reflectivity_scan_spec}, the points are not completely symmetric owing to the small variation in the circles' center position. We then fitted the analytical lower polariton dispersion to the measured points with a Rabi splitting of $\hbar\Omega_{\mathrm{Rabi}} = 11$\,meV, cavity energy $E_c = 1.612$\, meV, and exciton energy $E_x = 1.6088$\,meV \cite{Deng2010}. These values are in reasonable agreement with those reported earlier for identical  samples.

\begin{figure}
 \includegraphics[width=0.47\textwidth]{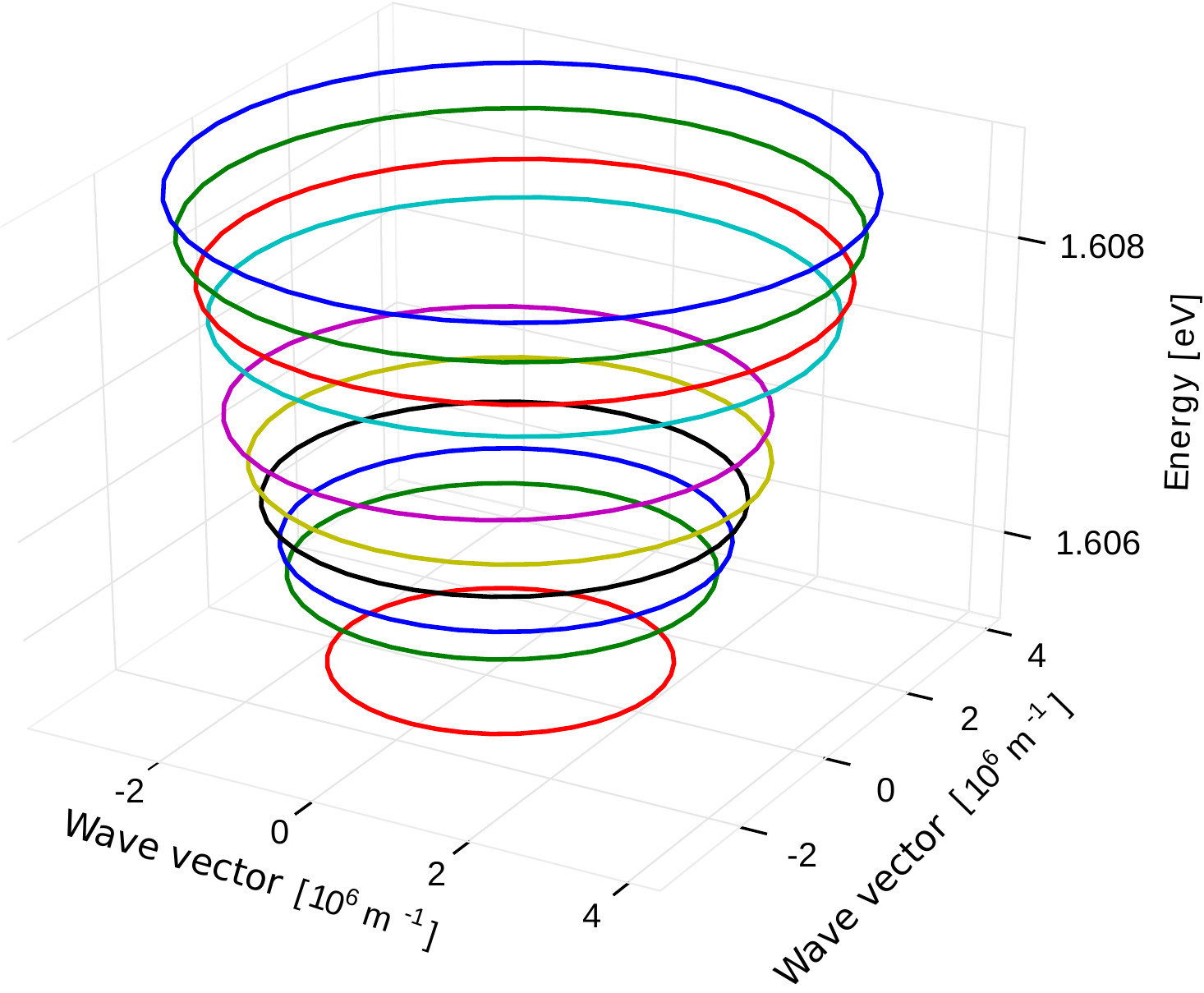}
 \caption{(color online) Three-dimensional representation of the measured polariton dispersion.}
 \label{fig:ple-3d}
\end{figure}

Next, we conducted a photoluminescence excitation (PLE) measurement. Again, the wavelength of the laser is kept constant at $770.2$\,nm, while the excitation momentum is scanned. However, instead of detecting the reflected power, we measure the luminescence intensity at the center of the dispersion curve. For momentum selection on the detection side, we used a multimode fiber of diameter $62.5\,\mu$m, (F1 in Fig.~\ref{fig:exp_setup}), and fed the collected light into a spectrometer.

Resonance behavior is apparent in Fig.~\ref{fig:ple-combined}, where we show the luminescence intensities as a function of the excitation momentum. The intensities are taken from spectra integrated over a range of 3 nm. The range was chosen such that
that it suppresses the excitation laser's contribution, which, however, is still visible at normal incidence, and high momenta. Further suppression of the scattered laser light can be achieved by measuring in orthogonal polarization with respect to the excitation.

In comparison with the reflectivity dip in Fig.~\ref{fig:reflectivity_scan}, the PLE intensities in Fig.~\ref{fig:ple-combined} give a much better defined resonance, and we can easily fit a circle to the maxima. Repeating this procedure for various wavelengths, we get a three-dimensional representation of the polariton dispersion, as depicted in Fig.~\ref{fig:ple-3d}.

We would like to point out that our setup has tremendous advantages in terms of speed. Since the required phase masks can efficiently be calculated in the video card of a computer, and the amplitude masks do not have to be calculated, the displayed phases or amplitudes can be re-defined in 1/60 or 1/100 of a second, respectively. These numbers are determined by the maximum refresh rate of the SLMs. The two-dimensional scans in Fig.~\ref{fig:reflectivity_scan} (reflectivity) and in Fig.~\ref{fig:ple-combined} (PLE), can be acquired in about 40 seconds.

In conclusion, we introduced a versatile beam-steering system that can be used to easily access the complete polariton phase-space, even if an excitation scheme requires multiple momenta and energies. This approach should make it possible to easily, and simultaneously excite non-degenerate polariton states. Due to the technical difficulties mentioned in the introduction, non-degenerate excitation of polaritons is a largely unexplored field. Beyond its fundamental importance, it could be useful for the generation of entangled photons \cite{Portolan2009,Liew2011}. We would also like to mention that the scheme that we outlined above can also be used for studying polaritons in confined geometries, either in wires, or in micropillars.

We acknowledge partial funding of this work by the Austrian Science Fund (FWF), Project P22979-N16. We are also indebted to D. Snoke for providing us with samples at the initial stages of our work.


\begin{thebibliography}{99}
\bibitem{note}The focal spot size and angular resolution are determined by several factors: a short focal length on the SLM would increase the angular resolution at the expense of the final spot size on the sample. At the same time, the focal length defined on the SLM has to be long enough, so that the various diffraction orders can be separated.
\end{thebibliography}

\end{document}